\documentclass[12pt]{article}
\usepackage{graphicx}
\usepackage{amsmath}
\usepackage{amssymb}
\usepackage{ulem}
\usepackage{enumerate}
\usepackage{cite}
\usepackage{latexsym}
\begin{document}
\begin{center}
{\bf {\large{Nilpotent Symmetries and Conserved Charges: BRST-Quantized Version of the 4D Abelian 2-Form Gauge Theory }}}

\vskip 2.5cm

{\sf  R. P. Malik$^{(a,b)}$}\\
$^{(a)}$ {\it Physics Department, Institute of Science,}\\
{\it Banaras Hindu University, Varanasi - 221 005, India}\\

\vskip 0.1cm

$^{(b)}$ {\it DST Centre for Interdisciplinary Mathematical Sciences,}\\
{\it Institute of Science, Banaras Hindu University, Varanasi - 221 005, India}\\
{\small {\sf {e-mails: rpmalik1995@gmail.com; malik@bhu.ac.in}}}
\end{center}

\vskip 1.5 cm

\noindent
{\bf Abstract:} We focus on the  Becchi-Rouet-Stora-Tyutin (BRST) quantized version of the four (3 + 1)-dimensional (4D) free Abelian 2-form
gauge theory and (i) discuss {\it its} infinitesimal, continuous and off-shell nilpotent (anti-)BRST and (anti-)co-BRST
 symmetry  transformations, and (ii) derive the corresponding Noether conserved (anti-)BRST and (anti-)co-BRST charges. We demonstrate that 
 {\it these} charges are {\it not} invariant under the {\it above} nilpotent symmetry transformations. We systematically derive the {\it modified} versions of the
above Noether charges which are found to be {\it invariant} under the above symmetry transformations. As a consequence,  {\it both} types of 
the above charges anticommute with each-other which proves their linear independence. We also derive the (anti-)BRST and 
(anti-)co-BRST invariant Curci-Ferrari (CF) type restrictions from the algebraic structures of the Noether conserved charges.
In all the above proofs and derivations, we take into account the deep mathematical relationship that exists between the infinitesimal and continuous symmetry
transformations and their generators as the Noether conserved charges.



\newpage

\section {Introduction}

The standard model of elementary particle physics (SMEPP) is one of the most successful theories in  the realm of 
theoretical high energy physics because there is  stunning degree of agreements between theory and experiments for some 
of the well-known physical processes
that are allowed by the basic selection rules of the SMEPP. However, the
conclusive experimental evidence of the rest masses of the neutrinos is one of the biggest blows to the very foundation of the theoretical aspects
of SMEPP which is based on the {\it interacting} non-Abelian 1-form (i.e. $p = 1$) gauge theories. Thus, it has been a challenge for the theoretical 
high energy physicists to go beyond the domain of 
validity of the SMEPP in a consistent manner. The idea behind the (super)string theories 
(see, e.g. [1-3] and references therein) is one of 
the most promising theoretical developments  in this direction. The quantum excitations of the (super)strings,  however, lead to the existence of the higher
$p$-form (i.e. $p = 2, 3... $) basic fields which imply that the theoretical reach of 
the (super)strings goes much beyond the validity of SMEPP in a very subtle (but convincing) manner.

Against the backdrop of the above paragraph, it is clear that the study of the higher $p$-form (i.e. $p = 2, 3... $) gauge theories is important because
of {\it their} relevance to the (super)string theories. The central purpose of our present endeavor is to focus on a few 
theoretical aspects of the 4D free Abelian 2-form gauge theory within the framework of the 
Becchi-Rouet-Stora-Tyutin (BRST) formalism [4-7]. In particular, we concentrate on the infinitesimal, continuous and off-shell nilpotent versions of the
(anti-)BRST and (anti-)co-BRST symmetry transformations, derive the 
corresponding Noether conserved (anti-)BRST and (anti-)co-BRST charges and study some of {\it their} (i) key
properties, and (ii) algebraic structure. The study of the {\it latter} (in the terminology of the symmetry transformation operators and their 
generators as the Noether conserved charges) leads to the derivations of the Curci-Ferrari (CF) type restrictions on our theory (which 
are found to be (anti-) BRST and (anti-)co-BRST invariant quantities).
The existence of the (non-)trivial CF-type restrictions\footnote{It has been shown in our earlier works [8,9]
on the non-Abelian 1-form,  Abelian 2-form and 3-form gauge theories  that the {\it non-trivial} CF-type
restrictions (responsible for the absolute anticommutativity of the BRST and anti-BRST symmetry operators) 
are connected with the geometrical objects called gerbes.}
is one of the hallmarks of the BRST-quantized gauge theories.

One of the highlights of our present endeavor is the observation that the Noether conserved (anti-)BRST and 
(anti-)co-BRST charges are {\it not} invariant under the infinitesimal, continuous and off-shell nilpotent 
(anti-)BRST and (anti-)co-BRST symmetry transportations, respectively. Thus, even though these Noether charges
are conserved (i.e. invariant w.r.t. the time-evolution of our BRST-quantized
field-theoretic system), they are {\it not} physical from the point of view of the symmetry considerations. We have systematically (see, e.g. [10])
derived the {\it modified} versions of these conserved Noether charges that are found to be {\it invariant} under the {\it above} nilpotent  transformations.  Hence, these {\it modified} versions of the conserved\footnote{We have exploited {\it only} the theoretical strengths of (i) the Gauss divergence
theorem, and (ii) the Euler-Lagrange (EL) equations of motion (EoM), in the derivation of the {\it modified} versions of the 
(anti-) BRST and (anti-)co-BRST charges. Hence, the {\it modified} versions of
charges are {\it conserved} quantities  as are the Noether (anti-BRST and (anti-)co-BRST charges (according to the basic tenets of Noether's theorem).} 
(anti-)BRST and (anti-)co-BRST charges are {\it physical} (as far as the
symmetry considerations are concerned). We have demonstrated that the Noether conserved charges and the {\it modified} versions of the conserved charges
anticommute with each-other. Hence, these sets of charges are linearly independent of each-other. We have 
{\it also} established that {\it both} types of charges
have their own identity and they play decisive  roles within the framework of BRST formalism. Whereas the 
Noether conserved charges are the generators for the off-shell nilpotent transformations, the {\it modified} versions 
of the conserved charges are {\it physical} and they play a crucial role in the
discussion on physicality criteria.

The theoretical contents of our present investigation are organized as follows. In the next section, we recapitulate the bare essentials of 
the off-shell nilpotent (anti-)BRST transformations for the coupled (but equivalent) Lagrangian densities and 
briefly mention about the Noether conserved (anti-)BRST charges. 
 Our section three deals with a brief synopsis of the (anti-)co-BRST symmetry transformations and conserved (anti-)co-BRST charges.
 The subject matter of our section four is concerned with the systematic derivations 
of the CF-type restrictions (on our BRST-quantized theory). 
Finally, in our section five, we summarize our key
results and point out the future perspective of our present investigation.



\section{Preliminaries: Nilpotent (Anti-)BRST Symmetries}


For our 4D BRST-quantized\footnote{ We choose the flat 4D background  Minkowskian spacetime
metric tensor  $\eta_{\mu\nu} $ = diag (+1, -1, -1, -1) so that the dot
product between two non-null vectors $S_\mu$ and $T_\mu$ is defined as : $S\cdot T = \eta_{\mu\nu} S^\mu T^\nu = S_0 T_0 - S_i T_i $ 
where (i) the Greek indices $\mu, \nu, \sigma... = 0, 1, 2, 3$ correspond to the time and space directions, 
(ii) the Latin indices $i, j, k...= 1, 2, 3$ stand for the {\it space} directions {\it only}, and (iii) the repeated indices are 
{\it always}  summed over.
 In the entire body of our text,
we adopt the convention of the left-derivative w.r.t. the fermionic fields at {\it appropriate} places. We follow 
the convention: $(\delta B_{\mu\nu}/\delta B_{\rho\sigma}) = \frac{1}{2!}\, (\delta^\rho_\mu \delta^\sigma_\nu - \delta^\rho_\nu \delta^\sigma_\mu) $
for the variation/differentiation w.r.t. the gauge field $B_{\mu\nu}$. 
The symbols used to denote the
off-shell nilpotent  (anti-)BRST and (anti-)co-BRST transformation operators are $s_{(a)b}$ and $s_{(a)d}$ and corresponding Noether 
conserved (anti-)BRST and (anti-)co-BRST charges carry the notations $Q_{(a)b}$ and $Q_{(a)d}$, respectively. While using the Gauss divergence 
theorem, we assume that there is no {\it non-trivial} topology at the boundary.} version of the free Abelian 2-form gauge theory,
the (anti-) BRST invariant 
Lagrangian  densities ${\cal L}_{(\bar B)} $ and ${\cal L}_{(B)} $, respectively, are (see, e.g. [11,12]) 
\begin{eqnarray}\label{1}
{\cal L}_{(\bar B)} &=& \dfrac{1}{12}\, H^{\mu\nu\sigma}\,H_{\mu\nu\sigma} 
- \bar B^\mu \Big [\partial^\nu B_{\nu\mu} + \dfrac{1}{2} \, \partial_\mu \phi \Big ]
- \dfrac{1}{2} \, \bar B^\mu\, \bar B_\mu - \Big (\partial^\mu \bar C^\nu - \partial^\nu \bar C^\mu \Big )\, \partial_\mu C_\nu \nonumber\\
 &-& \dfrac{1}{2}\,
\partial^\mu \bar \beta \,\partial_\mu \beta 
- \dfrac{1}{2} \, \Big [(\partial \cdot C) - \frac{1}{4} \, \lambda \Big ] \, \rho 
- \dfrac{1}{2} \, \Big [(\partial \cdot \bar C) + \frac{1}{4} \, \rho \Big ]\, \lambda, \nonumber\\
{\cal L}_{(B)} &=& \dfrac{1}{12}\, H^{\mu\nu\sigma}\,H_{\mu\nu\sigma} + B^\mu \Big [\partial^\nu B_{\nu\mu} - \dfrac{1}{2} \, \partial_\mu \phi \Big ]
- \dfrac{1}{2} \, B^\mu\, B_\mu - \Big (\partial^\mu \bar C^\nu - \partial^\nu \bar C^\mu \Big )\, \partial_\mu C_\nu \nonumber\\
 &-& \dfrac{1}{2}\,
\partial^\mu \bar \beta \,\partial_\mu \beta 
- \dfrac{1}{2} \, \Big [(\partial \cdot C) - \frac{1}{4} \, \lambda \Big ] \, \rho 
- \dfrac{1}{2} \, \Big [(\partial \cdot \bar C) + \frac{1}{4} \, \rho \Big ]\, \lambda, 
\end{eqnarray}
where the kinetic term (i.e. $ \frac{1}{12}\, H^{\mu\nu\sigma}\,H_{\mu\nu\sigma} $) for the 
Abelian 2-form [i.e. $B^{(2)} = \frac{1}{2!}\, B_{\mu\nu}\, (d x^\mu \wedge d x^\nu) $] gauge field $B_{\mu\nu}$
is defined through the totally antisymmetric
field-strength tensor $H_{\mu\nu\sigma} = \partial_\mu B_{\nu\sigma} + \partial_\nu B_{\sigma\mu} + \partial_\sigma B_{\mu\nu} $.
 The above Lagrangian densities are called as the {\it coupled}
Lagrangian densities because the {\it bosonic} Nakanishi-Lautrup auxiliary fields $B_\mu$ and  $\bar B_\mu$  are
{\it not} independent as they are required to satisfy the CF-type restriction:
$B_\mu + \bar B_\mu + \partial_\mu \phi = 0$
 where the scalar field $\phi$ is introduced into our theory because of the stage-one reducibility 
 of the gauge field $B_{\mu\nu}$. 
As far as the Faddeev-Popov (FP) ghost part of the Lagrangian  density is concerned, we note that
the fermionic (i.e. $C_{\mu}^{2}=\bar{C}_{\mu}^{2}=0, C_{\mu} \bar{C}_{\nu}+\bar{C}_{\nu} C_{\mu}=0, C_{\mu} C_{\nu}+C_{\nu} C_{\mu}=$ $0, 
\bar{C}_{\mu} \bar{C}_{\nu}+\bar{C}_{\nu} \bar{C}_{\mu}=0$) Lorentz vector (anti-)ghost fields $(\bar C_\mu)C_\mu$ carry the
ghost numbers $(-1)+1$ and  the (anti-)ghost fields $(\bar{\beta}) \beta$ are  the Lorentz scalar 
{\it bosonic}  (i.e. $\beta^2 \neq 0, \; \bar \beta^2 \neq 0, \; \beta\, \bar \beta = \bar \beta \, \beta$)
(anti-)ghost fields  which are endowed with the 
ghost numbers $(-2)+2$, respectively. In fact,  the {\it latter} fields  are the ghost-for-ghost fields. The 
fermionic (i.e. $\rho^2 = \lambda^2 = 0, \; \rho\,\lambda  + \lambda\, \rho = 0 $) auxiliary (anti-)ghost fields:
$\rho = -\, 2\, (\partial \cdot \bar C), \;  \lambda = +\, 2\, (\partial \cdot  C)$ carry the ghost numbers (-1)+1, respectively. These
(anti-)ghost fields are invoked into our BRST-quantized theory  to maintain the sanctity of the {\it unitarity} at any arbitrary order
of perturbative computations for a given physical process that is allowed by the theory.

It is straightforward to check that, under the following 
infinitesimal, continuous and off-shell nilpotent (i.e. $s_{(a)b}^2 = 0 $)
(anti-)BRST symmetry transformations ($s_{(a)b}$), namely;
\begin{eqnarray}\label{2}
&& s_{ab} B_{\mu\nu} = -\, (\partial_\mu \bar C_\nu - \partial_\nu \bar C_\mu), \quad s_{ab} \bar C_\mu = -\, \partial_\mu \bar \beta, 
\quad s_{ab}  C_\mu = \bar B_\mu, 
\quad s_{ab} \phi = +\, \rho, \nonumber\\
&&s_{ab}  \beta = -\, \lambda, \qquad \;s_{ab}  B_\mu = -\, \partial_\mu \rho, \qquad \;
s_{ab} \big [\lambda, \;\rho, \; \bar \beta,  \; \bar B_\mu, \; H_{\mu\nu\sigma} \big ] = 0, \nonumber\\
&& s_b B_{\mu\nu} = -\, (\partial_\mu C_\nu - \partial_\nu C_\mu), \qquad s_b C_\mu = -\, \partial_\mu \beta, \qquad s_b \bar C_\mu = B_\mu, 
\qquad s_b \phi = \lambda, \nonumber\\
&&s_b \bar \beta = -\, \rho, \qquad \; s_b \bar B_\mu = -\, \partial_\mu \lambda, \qquad \;
s_b \big [\lambda, \;\rho, \;\beta, \; B_\mu, \; H_{\mu\nu\sigma} \big ] = 0,
\end{eqnarray}
the Lagrangian  densities ${\cal L}_{(\bar B)}$ and ${\cal L}_{(B)}$ transform to the {\it total}
spacetime derivatives, namely;
\begin{eqnarray}\label{3}
s_{ab} \,{\cal L}_{(\bar B)} &=& +\, \partial_\mu \Big [ (\partial^\mu \bar C^\nu - \partial^\nu \bar C^\mu)\, \bar B_\nu 
+ \dfrac{1}{2}\, \lambda\, \partial^\mu \bar \beta - 
\dfrac{1}{2}\, \rho\, \bar B^\mu \Big ], \nonumber\\
s_b\, {\cal L}_{(B)} &=& -\, \partial_\mu \Big [ (\partial^\mu C^\nu - \partial^\nu C^\mu)\, B_\nu + \dfrac{1}{2}\, \lambda\, B^\mu - 
\dfrac{1}{2}\, \rho\, \partial^\mu\, \beta \Big ], 
\end{eqnarray}
thereby rendering the action integrals  $S_1 = \int d^4 x \,{\cal L}_{(\bar B)} $ and $S_2 = \int d^4 x \,{\cal L}_{(B)} $
invariant (i.e. $ s_{ab}\, S_1 = 0 , \; s_b \,S_2 = 0$)  due to the validity of the
 Gauss divergence theorem. Applying the standard mathematical formula for the
Noether theorem, we obtain the expressions for the  Noether conserved (anti-)BRST currents 
which lead to the derivations of the Noether conserved (anti-)BRST charges  
(i.e. $Q_{(a)b}$) as follows (see, e.g. [13] for details): 
\begin{eqnarray}\label{4}
Q_{ab} &=& \int d^3 x\,\, \Big [\big (\partial^0 C^i - \partial^i C^0 \big ) \, \partial_i \bar \beta + \big (\partial^0 \bar C^i 
- \partial^i \bar C^0 \big ) \, \bar B_i \nonumber\\
&+& \frac{1}{2}\, \lambda \, \dot {\bar \beta} - \frac{1}{2}\, \rho \, \bar B^0 - \dfrac{1}{2}\, H^{0ij} \,
(\partial_i \bar C_j - \partial_j \bar C_i)\Big ], \nonumber\\
Q_{b} &=& - \int d^3 x\,\,
\Big [\big (\partial^0 \bar C^i - \partial^i \bar C^0 \big ) \, \partial_i  \beta + \big (\partial^0  C^i 
- \partial^i  C^0\big ) \,  B_i \nonumber\\
&-& \frac{1}{2}\, \rho \, \dot  \beta + \frac{1}{2}\, \lambda \,  B^0 + \dfrac{1}{2}\, H^{0ij} \,
(\partial_i C_j - \partial_j  C_i) \Big ].
\end{eqnarray}
These Noether conserved (anti-)BRST charges are found to be the generators for the 
infinitesimal, continuous and off-shell nilpotent (anti-)BRST 
transformations (2) through the following explicit mathematical relationship (see, e.g. [13] for details)
\begin{eqnarray}\label{5}
 s_{(a)b} \, \Phi (\vec{x}, t)\, = \,-\, i\, \Big [\Phi (\vec{x}, t), \; \; Q_{(a)b} \Big ]_{(\pm)},
\end{eqnarray}
where the subscripts $(\pm)$ on the square bracket
denote the (anti)commutator for the generic dynamical field $\Phi$ of our theory [cf. Eq. (1)]  being fermionic/bosonic in nature.

We conclude our present section  with a few crucial remarks. First of all, we note that  the (anti-)BRST invariant CF-type restriction
(i.e. $B_\mu + \bar B_\mu + \partial_\mu \phi = 0$) is responsible for the absolute anticommutativity property (i.e. $\{ s_b, \; s_{ab} \} = 0$) 
between the BRST and
anti-BRST symmetry transformation operators $s_{(a)b} $ because we observe that: 
\begin{eqnarray}\label{6}
 \big \{s_b , \; s_{ab} \big \}\, B_{\mu\nu} = -\, \partial_\mu (B_\nu + \bar B_\nu) + \partial_\nu (B_\mu + \bar B_\mu).
\end{eqnarray}
The above equation turns out to be {\it zero}  
on the space of quantum fields where the {\it above} restriction is satisfied. For the {\it rest} of the quantum fields of  
our theory [cf. Eq. (1)], the absolute anticommutativity property (i.e. $\{ s_b, \; s_{ab} \} = 0$) is {\it trivially} satisfied.
Second, we note that the Noether 
conserved (anti-)BRST charges $Q_{(a)b} $ are {\it not} invariant under the infinitesimal, continuous and {\it off-shell}
nilpotent(anti-)BRST transformation operators (2) because we observe that the following expressions are true, namely;
\begin{eqnarray}\label{7}
s_{ab} \,Q_{ab} &=& \int d^3 x\, \Big [\big (\partial^0 \bar B^i - \partial^i \bar B^0 \big ) \, \partial_i \bar \beta \Big ] \neq 0, \nonumber\\
s_b \,Q_{b} &=& - \int d^3 x\, \Big [\big (\partial^0  B^i - \partial^i  B^0 \big ) \, \partial_i  \beta \Big ] \neq 0.
\end{eqnarray}
Hence, these Noether conserved (anti-)BRST charges $Q_{(a)b} $ are {\it not} physical in the true sense of the word (within the framework of BRST formalism).
Thus, we have to find out the consistently {\it modified} versions of 
the above Noether conserved charges which remain {\it invariant} under the infinitesimal, continuous
and {\it off-shell} nilpotent (anti-)BRST transformations.  Such an exercise has already been performed in our earlier work [13] where we have obtained the 
following expressions for the {\it modified} versions of the (anti-)BRST charges $Q_{(A)B}$ from the 
Noether conserved charges $Q_{(a)b}$ [cf. Eq. (4)], namely;
\begin{eqnarray}\label{8}
Q_{ab} \longrightarrow Q_{AB} &=& \int d^3 x\,\,
\Big [\big (\partial^0 \bar  C^i 
- \partial^i  \bar C^0 \big ) \,  \bar B_i -\, \big (\partial^0 \bar B^i - \partial^i \bar B^0 \big )\, \bar C_i \nonumber\\
&-&
\big (\partial^0  C^i - \partial^i C^0 \big ) \, \partial_i  \bar \beta  
+ \frac{1}{2}\, \lambda\, \dot  {\bar \beta} - \frac{1}{2}\, \rho \,  \bar B^0 - \dot \lambda\, \bar \beta   \Big ], \nonumber\\
Q_b \longrightarrow Q_{B} &=& \int d^3 x\,\,
\Big [ \big (\partial^0 B^i - \partial^i B^0 \big )\, C_i +
\big (\partial^0 \bar C^i - \partial^i \bar C^0 \big ) \, \partial_i  \beta \nonumber\\ 
&-& \big (\partial^0  C^i - \partial^i  C^0\big ) \,  B_i 
- \frac{1}{2}\, \rho \, \dot  \beta + \frac{1}{2}\, \lambda \,  B^0 - \dot \rho\, \beta   \Big ],
\end{eqnarray}
which have been derived [from  the Noether conserved (anti-)BRST charges $Q_{(a)b}$]
 by exploiting the theoretical strength of (i) the 
partial integration along with the Gauss divergence theorem,
and (ii) the appropriate EL-EoM at suitable places. Hence, the above {\it modified} versions 
of the (anti-)BRST charges are {\it conserved} quantities as are the Noether conserved (anti-)BRST charges.
In view of the relationship (5), we have the following:
\begin{eqnarray}\label{9}
s_{ab} \,Q_{AB} = -\, i\, \big \{Q_{AB}, \; Q_{ab} \big \} = 0, \qquad \quad  s_{b} \,Q_{B} = -\, i\, \big \{Q_{B}, \; Q_{b} \big \} = 0.
\end{eqnarray}
In other words, we have (i) a set of {\it two} conserved BRST charges (i.e. $Q_b $ and $Q_B $)
which are linearly independent of each-other, and (ii) the corresponding set of {\it two} conserved anti-BRST charges (i.e. $Q_{ab} $ and $Q_{AB} $) 
which are {\it also} linearly independent of each-other. 
Whereas the Noether conserved (anti-)BRST charges $Q_{(a)b} $ are the generators (see, e.g. [13] for details)
for the (anti-)BRST symmetry transformations (2), the {\it modified} versions of the (anti-)BRST charges $Q_{(A)B} $ are useful 
in the physicality criteria (see, e.g. [13] for details). 



\section{Nilpotent (Anti-)co-BRST Symmetries}

In the specific {\it four} dimensions of the flat Minkowskian spacetime, the {\it kinetic} term 
for the gauge field $B_{\mu\nu}$ can be linearized by invoking a 
new set of {\it bosonic} Nakanishi-Lautrup type auxiliary fields ($\bar {\cal B}_\mu, {\cal B}_\mu $) along with a
derivative on the pseudo-scalar field ($\widetilde \phi $) in a linearly independent fashion. The resulting {\it coupled} Lagrangian
densities\footnote{The 4D totally antisymmetric 
Levi-Civita tensor $\varepsilon_{\mu\nu\sigma\rho}$ is chosen such that:  $\varepsilon_{0123} 
= +1 = -\, \varepsilon^{0123}$ and, when two of them are contracted in a specific fashion,
 they satisfy the standard relationships: $\varepsilon_{\mu\nu\eta\kappa} \varepsilon^{\mu\nu\eta\kappa}= - \,4!,\;\;$
$\varepsilon_{\mu\nu\eta\kappa}\, \varepsilon^{\mu\nu\eta\rho}= - \,3! \;\;\delta^\rho_\kappa, \; \varepsilon_{\mu\nu\eta\kappa} \,
\varepsilon^{\mu\nu\sigma\rho}= - \,2!\;\;
\big(\delta^\sigma_\eta \delta^\rho_\kappa - \delta^\sigma_\kappa \delta^\rho_\eta\big)$, etc.} 
are [11,12]
\begin{eqnarray}\label{10}
&&{\cal L}_{(\bar {\cal B}, \bar B)} = 
 \dfrac{1}{2} \bar {\cal B}_\mu \bar {\cal B}^\mu + \bar {\cal B}^\mu \Big (\dfrac{1}{2}
\varepsilon_{\mu\nu\sigma\rho} \, \partial^\nu B^{\sigma\rho} + \dfrac{1}{2}\,\partial_\mu \widetilde \varphi \Big)
- \bar B^\mu \Big [\partial^\nu B_{\nu\mu} + \dfrac{1}{2} \, \partial_\mu \phi \Big ]
- \dfrac{1}{2} \, \bar B^\mu\, \bar B_\mu \nonumber\\
&&- \Big (\partial^\mu \bar C^\nu - \partial^\nu \bar C^\mu \Big )\, \partial_\mu C_\nu 
 - \dfrac{1}{2}\,
\partial^\mu \bar \beta \,\partial_\mu \beta 
- \dfrac{1}{2} \, \Big [(\partial \cdot C) - \frac{1}{4} \, \lambda \Big ] \, \rho 
- \dfrac{1}{2} \, \Big [(\partial \cdot \bar C) + \frac{1}{4} \, \rho \Big ]\, \lambda, \nonumber\\
&& {\cal L}_{({\cal B}, B)} = 
\dfrac{1}{2}{\cal B}_\mu {\cal B}^\mu - {\cal B}^\mu \Big (\dfrac{1}{2}
\varepsilon_{\mu\nu\sigma\rho} \, \partial^\nu B^{\sigma\rho} - \dfrac{1}{2}\,\partial_\mu \widetilde \varphi \Big )
+ B^\mu \Big [\partial^\nu B_{\nu\mu} - \dfrac{1}{2} \, \partial_\mu \phi \Big ]
- \dfrac{1}{2} \, B^\mu\, B_\mu \nonumber\\
&& - \Big (\partial^\mu \bar C^\nu - \partial^\nu \bar C^\mu \Big )\, \partial_\mu C_\nu 
 - \dfrac{1}{2}\,
\partial^\mu \bar \beta\, \partial_\mu \beta 
- \dfrac{1}{2}  \Big [(\partial \cdot C) - \frac{1}{4} \, \lambda \Big ] \, \rho 
- \dfrac{1}{2}  \Big [(\partial \cdot \bar C) + \frac{1}{4} \, \rho \Big ]\, \lambda, 
\end{eqnarray}
where the Nakanishi-Lautrup auxiliary fields $\bar {\cal B}_\mu $ and ${\cal B}_\mu $ are {\it not} independent fields because they are required to
satisfy the CF-type restriction: ${\cal B}_\mu + \bar {\cal B}_\mu + \partial_\mu \widetilde \phi = 0 $. It is straightforward to note that under
the following infinitesimal, continuous and off-shell nilpotent (i.e. $s_{(a)d}^2 = 0 $)
(anti-)dual BRST [i.e. (anti-)co-BRST] transformations\footnote{The total gauge-fixing term 
for the gauge field (owing its origin to the co-exterior derivative 
of differential geometry [14-16]) remains {\it invariant} under the (anti-)co-BRST 
symmetry transformations.  On the other hand, 
the total kinetic term for the gauge field (owing its origin to the exterior derivative [14-16]) stays {\it unchanged} under the (anti-)BRST 
symmetry transformations (see, e.g. [11,12] for details).}
 $s_{(a)d}$ (see, e.g. [11,12]) 
\begin{eqnarray}\label{11}
&&  s_{ad} B_{\mu\nu} = - \varepsilon_{\mu\nu\sigma\rho} \,\partial^\sigma  C^\rho, \qquad 
s_{ad} \bar C_\mu = {\bar{\cal  B}}_\mu, \qquad s_{ad} C_\mu  =  \partial_\mu  \beta, \qquad s_{ad} \widetilde \varphi = -\,\lambda, \nonumber\\
&& s_{ad}  \bar\beta =  \rho, \qquad  \quad s_{ad} {\cal{ B}}_\mu =   \partial_\mu \lambda,\qquad
s_{ad} \big [\rho, \lambda, \beta, B_\mu, {\cal{\bar B}}_\mu,
\bar B_\mu, (\partial^\nu B_{\nu\mu}) \big ] = 0,
\nonumber\\
&&  s_d B_{\mu\nu} = - \varepsilon_{\mu\nu\sigma\rho} \,\partial^\sigma \bar C^\rho, \qquad 
s_d  C_\mu = {\cal B}_\mu, \qquad s_d \bar C_\mu  = - \partial_\mu \bar \beta, \qquad s_{d} \widetilde \varphi = -\,\rho,\nonumber\\
&& s_d  \beta = - \lambda, \qquad s_d {\bar {\cal B}}_\mu =   \partial_\mu \rho, \qquad
s_d \big [\rho, \lambda, \bar\beta, B_\mu, {\cal{ B}}_\mu,\bar B_\mu, (\partial^\nu B_{\nu\mu}) \big] = 0, 
\end{eqnarray}
the Lagrangian densities ${\cal L}_{(\bar {\cal B}, \bar B)} $ and ${\cal L}_{({\cal B}, B)} $ transform to the total spacetime derivatives as 
\begin{eqnarray}\label{12}
s_{ad} {\cal L}_{( {\cal{\bar B}}, \bar B)} &=& - \,\partial_\mu \, \Big [ 
 \big(\partial^\mu C^\nu-\partial^\nu C^\mu \big)\, \bar {\cal B}_\nu + \dfrac{1}{2}\,\lambda \,\bar {\cal B}^\mu   + 
\dfrac{1}{2}\, \rho\, \partial^\mu \beta \, \bigg], \nonumber\\
s_d {\cal L}_{( {\cal B}, B)} &=& +  \, \partial_\mu \, 
\bigg[   
\big(\partial^\mu {\bar C}^\nu -\partial^\nu \bar C^\mu \big)\,{\cal B}_\nu  + \dfrac{1}{2}\,\lambda \, \partial^\mu \bar\beta 
- \dfrac{1}{2}\,  \rho \,{\cal B}^\mu \bigg], 
\end{eqnarray}
thereby rendering the action integrals $\widetilde S_1 = \int d^4 x\, {\cal L}_{({\cal{\bar B}}, \bar B) } $ and 
$ \widetilde S_2 = \int d^4 x\, {\cal L}_{({\cal{B}}, B)} $
invariant (i.e. $s_{ad} \widetilde S_1 = 0, \; s_d \widetilde S_2 = 0 $) due to the validity of Gauss's divergence theorem.

The observations in equation (12) imply that we can exploit the theoretical potential of the Noether theorem to derive the Noether conserved currents
which lead to the derivations of the Noether conserved charges. We have already performed this exercise in our earlier work [11] on the 
BRST-quantized version of the 4D
St{\" u}ckelberg-modified {\it massive} Abelian 2-form gauge theory whose 
{\it massless} limit will lead to the expressions for the Noether conserved (anti-)co-BRST charges for our {\it present}  theory. In other words, 
we have the following precise expressions for the Noether conserved  (anti-)co-BRST charges $Q_{(a)d}$ for our 
4D BRST-quantized version of the {\it free} Abelian 2-form gauge theory, namely;
\begin{eqnarray}\label{13}
Q_{ad} &=& \int d^3 x\, \Big[\varepsilon^{0ijk}\, \bar B_i \, \big(\partial_j C_k \big)
+  \big(\partial^0 \bar C^i - \partial^i \bar C^0 \big)\, \partial_i \beta \nonumber\\
&-&  \big(\partial^0 C^i - \partial^i C^0 \big)\, \bar {\cal B}_i - \dfrac{1}{2}\, \rho  \, \partial^0 \beta 
- \dfrac{1}{2}\, \lambda \, \bar {\cal B}^0  \Big],
\nonumber\\
Q_d &=& \int d^3 x \Big[\big(\partial^0 C^i - \partial^\nu C^0 \big)\,\partial_i \bar \beta  -
\varepsilon^{0ijk}\,B_i\, \big(\partial_j \bar C_k \big)  
\nonumber\\
&+&   \big(\partial^0 \bar C^i - \partial^i \bar C^0 \big) {\cal B}_i + \dfrac{1}{2}\, \lambda \, \partial^0 \bar \beta
- \dfrac{1}{2} \rho\, {\cal B}^0 \Big],
\end{eqnarray}
which turn out to be the generators for the infinitesimal, continuous and off-shell nilpotent (anti-)co-BRST symmetry transformations (11)
if we use the mathematical relationship of equation (5) with the replacements: $s_{(a)b} \to  s_{(a)d}, \; Q_{(a)b} \to Q_{(a)d}$
and take into account the generic field $\Phi$ as the quantum fields that are present in the 
{\it coupled} Lagrangian densities (10). At this crucial juncture, 
we note that the Noether conserved (anti-)co-BRST charges are {\it not} invariant under the (anti-)co-BRST transformations (11) 
because we observe:
\begin{eqnarray}\label{14}
s_{ad} \,Q_{ad} = \int d^3 x\, \Big [\big (\partial^0 \bar {\cal B}^i - \partial^i \bar {\cal B}^0 \big ) \, \partial_i  \beta \Big ] \neq 0, \quad
s_d \,Q_{d} =  \int d^3 x\, \Big [\big (\partial^0 {\cal  B}^i - \partial^i  {\cal B}^0 \big ) \, \partial_i  \bar \beta \Big ] \neq 0.
\end{eqnarray}
In other words, the Noether conserved (anti-)co-BRST charges (even though invariant w.r.t. the time-evolution
of our BRST-quantized field-theoretic system) are  {\it not} physical\footnote{Our (i) present 4D BRST-quantized free Abelian 2-form gauge theory [12], and 
(ii) the St{\" u}ckelberg-modified {\it massive}
Abelian 2-form gauge theory [11] have been proven to be the field-theoretic examples for Hodge theory. In such theories, the physical quantities are 
{\it those} that remain {\it invariant} under the 
infinitesimal, continuous and off-shell nilpotent (anti-)BRST and (anti-)co-BRST transformations.} quantities from the point of view of 
the (anti-)co-BRST symmetry considerations.

Against the backdrop of the above paragraph, we derive the (anti-)co-BRST invariant (i.e. $s_{ad} Q_{AD} = 0, \; s_d Q_D = 0 $) versions of the
(anti-)co-BRST charges $Q_{(A)D}$ from their
counterparts  Noether conserved (anti-)co-BRST charges $Q_{(s)d}$ in a consistent manner where we exploit {\it only}
the Gauss divergence theorem and appropriate EL-EoM at suitable places. These theoretical exercises
 ensure that the {\it modified} (anti-)co-BRST charges $Q_{(A)D}$ are {\it conserved} quantities as per the rules of Noether's theorem. In this context,
 first of all, we focus on the {\it second} term (i.e. $ - \int d^3 x \, \varepsilon^{0ijk}\, B_i \,\partial_j \bar C_k$) of the  Noether 
 co-BRST charge $Q_d$ [cf. Eq. (13)]. Using the partial integration along with Gauss's divergence theorem, we obtain:
\begin{eqnarray}\label{15}
-\, \int d^3 x \;\varepsilon^{0ijk}\,B_i\, \big(\partial_j \bar C_k \big) = + \, \int d^3 x \; \varepsilon^{0ijk}\, (\partial_j B_i) \;\bar C_k
\equiv - \, \int d^3 x \;\varepsilon^{0kij}\, (\partial_i B_j) \;\bar C_k.
\end{eqnarray}
At this stage, using the EL-EoM: $\varepsilon^{0kij}\, \partial_i B_j = (\partial^0 {\cal B}^k - \partial^k {\cal B}^0)$ (see, e.g. [11]), we obtain 
\begin{eqnarray}\label{16}
 - \, \int d^3 x \,\varepsilon^{0ijk}\, (\partial_i B_j) \;\bar C_k = -\, \int d^3 x\, (\partial^0 {\cal B}^i - \partial^i {\cal B}^0)\, \bar C_i,
\end{eqnarray} 
which will be {\it present} in the expression for the {\it modified} version of the co-BRST charge $Q_D$. It is worthwhile to mention that in the process of
obtaining the equation (16) from equation (15), we have used the EL-EoM:  
$\varepsilon^{\mu\nu\sigma\rho}\, \partial_\sigma B_\rho = (\partial^\mu {\cal B}^\nu - \partial^\nu {\cal B}^\mu)$ which is derived from the
Lagrangian density ${\cal L}_{({\cal B}, B)} $ (see, e.g. [11]). Toward our main goal of obtaining the explicitly co-BRST invariant (i.e. $s_d\, Q_D = 0 $)
dual-BRST charge $Q_D$, we have to apply the co-BRST symmetry transformation on the r.h.s. of equation (16) because {\it this} term would be present, as 
claimed earlier, in the expression for $Q_D$. Accordingly, we have the following:
\begin{eqnarray}\label{17}
 s_d \Big [-\, \int d^3 x\, (\partial^0 {\cal B}^i - \partial^i {\cal B}^0)\, \bar C_i 
=  \int d^3 x\, (\partial^0 {\cal B}^i - \partial^i {\cal B}^0)\, \partial_i \bar \beta.
\end{eqnarray} 
At this stage, as per the rule proposed in our earlier work [10], we have to modify a specific term from the expression for the Noether conserved
co-BRST charge $Q_d$ such that when $s_d$ acts on  a part of that {\it modified} term, the ensuing outcome {\it must}  
cancel out with our result in equation (17).
Such a modified term from $Q_d$ [cf. Eq. (13)] is as follows
\begin{eqnarray}\label{18}
\int d^3 x \big(\partial^0 C^i - \partial^\nu C^0 \big)\,\partial_i \bar \beta = 2 \int d^3 x \big(\partial^0 C^i
 - \partial^\nu C^0 \big)\, \partial_i \bar \beta 
- \int d^3 x \big(\partial^0 C^i - \partial^\nu C^0 \big)\, \partial_i \bar \beta,  
\end{eqnarray}
which is nothing but the {\it first} integral on the r.h.s. of $Q_d$ [cf. Eq. (13)]. It is straightforward to note that if $s_d$ acts on the
{\it second} integral (with the inputs: $s_d C_\mu = {\cal B}_\mu, \; s_d \bar \beta = 0$) 
on the r.h.s. of the {\it above} equation, the ensuing result will cancel out with the result of equation (17). Hence, {\it this}
second integral of equation (18)
will be present in the explicit expression for $Q_D$. 
Thus far, we have already obtained {\it two} very important integrals of $Q_D$ as (i) the 
integral on the r.h.s. of equation (17),  and (ii) the {\it second} integral on the r.h.s. of equation (18). 
As far as the {\it first} integral on the r.h.s. of (18) is concerned, we apply the partial integration along with the 
Gauss divergence theorem to recast it in the following form:
\begin{eqnarray}\label{19}
2\, \int d^3 x \Big[\big(\partial^0 C^i
 - \partial^\nu C^0 \big)\, \Big ] \, (\partial_i \bar \beta)  = 
&-& 2\, \int d^3 x \Big[ \partial_i \big(\partial^0 C^i - \partial^\nu C^0 \big)\,\Big ]\, \bar \beta.  
\end{eqnarray}
Using the following EL-EoM 
w.r..t the anti-ghost field $\bar C_\mu$, namely;
\begin{eqnarray}\label{20}
 \partial_\mu \big (\partial^\mu C^\nu - \partial^\nu  C^\mu \big ) + \dfrac{1}{2} \, \partial^\nu \lambda = 0 
 \;\;\Longrightarrow \; \; \partial_i \big (\partial^0  C^i - \partial^i  C^0 \big ) = +\, \dfrac{1}{2}\, \dot \lambda,
\end{eqnarray}
we note that integral in (19) reduces to: $-\, \int d^3 x (\dot \lambda \, \bar \beta)$ which turns out to be a co-BRST invariant 
(i.e. $s_d [\dot \lambda \, \bar \beta] = 0 $) quantity. Hence, this integral will  {\it also} be 
present in the explicit expression for $Q_D$. We find that 
{\it rest} of all the integrals of the Noether co-BRST charge $Q_d$ [cf. Eq. (13)]
are co-BRST invariant quantities. As a consequence, they will be automatically present in the 
exact expression for $Q_D$. The {\it final} form of $Q_D$ is:
\begin{eqnarray}\label{21}
Q_d \to Q_D &=& \int d^3 x \Big[  \big(\partial^0 \bar C^i - \partial^i \bar C^0 \big)\, {\cal B}_i
- \big(\partial^0 C^i - \partial^\nu C^0 \big)\,\partial_i \bar \beta  - \big(\partial^0 {\cal B}^i - \partial^i {\cal B}^0 \big)\, \bar C_i 
  \nonumber\\
&+&    \dfrac{1}{2}\, \lambda \, \dot {\bar \beta}
- \dfrac{1}{2}\, \rho\, {\cal B}^0  - \dot \lambda\, \bar \beta \Big].
\end{eqnarray}
If we use the co-BRST symmetry transformations $s_d$ [cf. Eq. (11)], it is straightforward to check that $s_d\, Q_D = 0$. In view of equation (5),
we observe that $s_d Q_D = -\, i\, \{ Q_D, \; Q_d \} = 0 $.

We end this section with the remark that exactly {\it similar} kinds of exercises can be performed to obtain the anti-co-BRST invariant 
(i.e. $s_{ad} \,Q_{AD} = 0$) version of the anti-co-BRST charge $Q_{AD}$ (from the Noether conserved anti-co-BRST charge $Q_{ad}$) as:
\begin{eqnarray}\label{22}
Q_{ad} \to Q_{AD} &=& \int d^3 x \Big[ \big(\partial^0 \bar {\cal B}^i - \partial^i \bar {\cal B}^0 \big)\, C_i 
- \big(\partial^0  C^i - \partial^i  C^0 \big)\, \bar  {\cal B}_i
- \big(\partial^0 \bar C^i - \partial^\nu \bar C^0 \big)\,\partial_i  \beta   
  \nonumber\\
&-&    \dfrac{1}{2}\, \lambda \, \bar {\cal B}^0
- \dfrac{1}{2} \rho\, \dot \beta  + \dot \rho\,  \beta \Big].
\end{eqnarray}
In view of the key relationship in equation (5),
we observe that $s_{ad}\, Q_{AD} = -\, i\, \{ Q_{AD}, \; Q_{ad} \} = 0 $. In other words, we find that the 
set of Noether conserved (anti-)co-BRST charges $Q_{(a)d}$
and the {\it modified} versions of the 
conserved (anti-)co-BRST charges $Q_{(A)D}$ {\it  anticommute} with each-other. Hence, these {\it two}
types of conserved charges are 
linearly independent
of each-other and they have their own identity within the framework of BRST formalism.



\section{CF-Type Restrictions: Algebraic Structures}

The central purpose of our present section is to exploit the theoretical strength of the relationship (5) 
to focus on the algebraic structures amongst the
Noether conserved (anti-)BRST and (anti-)co-BRST charges 
and to establish that (i) the absolute anticommutativity between the Noether 
conserved BRST and anti-BRST charges is satisfied if and only if the CF-type restriction: $B_\mu + \bar B_\mu + \partial_\mu \phi = 0 $ is invoked in 
{\it its} proof.
On the other hand, the CF-type restriction: ${\cal B}_\mu + \bar {\cal B}_\mu + \partial_\mu \widetilde \phi = 0 $ is required in the proof of the
absolute anticommutativity between the Noether conserved co-BRST and anti-co-BRST charges.  In this context, first of all, we concentrate on the
relationship: $s_b\, Q_{ab} = -\, i\, \{Q_{ab}, \; Q_b \}$ [cf. Eq. (5)] and explicitly compute the l.h.s. (i.e. $s_b \,Q_{ab} $) by taking into account
(i) the off-shell nilpotent BRST symmetry transformations from equation (2), (ii) the explicit expression for $Q_{ab}$ from equation (4), (iii)
the partial integration along with the theoretical strength of the Gauss divergence theorem, and (iv) the appropriate EL-EoM at suitable places.
It is crystal clear that 
the operation of the BRST symmetry transformation operator $s_b$ 
[cf. Eq. (2)] on the Noether conserved anti-BRST charge $Q_{ab}$ [cf. Eq. (4)] leads to 
the following: 
\begin{eqnarray}\label{23}
s_b \,Q_{ab} &=& \int d^3 x\,\, \Big [\big (\partial^0 C^i - \partial^i C^0 \big ) \, \partial_i \rho + \big (\partial^0 \bar C^i 
- \partial^i \bar C^0 \big ) \, \partial_i \lambda + \big (\partial^0 B^i - \partial^i B^0 \big ) \, \bar B_i \nonumber\\
&+& \dfrac{1}{2}\, \lambda \, \dot \rho - \dfrac{1}{2}\, \rho \, \dot \lambda -  H^{0ij} \, \partial_i B_j\Big ] 
\equiv -\, i\, \Big \{Q_{ab}, \; Q_b \Big \}.
\end{eqnarray}
At this juncture of our discussion, we can exploit the theoretical strengths of (i) the partial integration along with the Gauss divergence theorem, and
(ii) the appropriate EL-EoM at suitable places. In this context, first of all, we 
exploit the beauty of the partial integration and Gauss's divergence theorem on the following {\it four} integrals of the above equation:
\begin{eqnarray}\label{24}
\int d^3 x\,\, \Big [\big (\partial^0 C^i - \partial^i C^0 \big ) \, \partial_i \rho + \big (\partial^0 \bar C^i 
- \partial^i \bar C^0 \big ) \, \partial_i \lambda 
+ \dfrac{1}{2}\, \lambda \, \dot \rho - \dfrac{1}{2}\, \rho \, \dot \lambda \Big ].
\end{eqnarray}
As a consequence, this equation gets converted into the following key result, namely;
\begin{eqnarray}\label{25}
\int d^3 x\,\, \Big [ -\, \partial_i \,\big (\partial^0 C^i - \partial^i C^0 \big ) \,  \rho -  \partial_i\, \big (\partial^0 \bar C^i 
- \partial^i \bar C^0 \big ) \,  \lambda 
+ \dfrac{1}{2}\, \lambda \, \dot \rho - \dfrac{1}{2}\, \rho \, \dot \lambda \Big ] = 0,
\end{eqnarray}
where we have used the EL-EoM of equation (20) and its counterpart, namely;
\begin{eqnarray}\label{26}
 \partial_\mu \big (\partial^\mu \bar C^\nu - \partial^\nu \bar  C^\mu \big ) - \dfrac{1}{2} \, \partial^\nu \rho = 0 
 \;\;\Longrightarrow \; \; \partial_i \big (\partial^0 \bar C^i - \partial^i  \bar C^0 \big ) = -\, \dfrac{1}{2}\, \dot \rho.
\end{eqnarray}
Now we deal with the {\it last} integral of the expression for  $s_b \, Q_{ab}$ [cf. Eq. (23)]. 
Using the partial integration along with the Gauss divergence theorem, we obtain the following
\begin{eqnarray}\label{27}
-\, \int d^3 x\,\,  H^{0ij} \, \partial_i B_j = +\, \int d^3 x\,\big (\partial_i H^{0ij} \big ) \,  B_j
\equiv -\, \int d^3 x\,\big (\partial^0 \bar B^i - \partial^i \bar B^0 \big ) \, \bar B_i,
\end{eqnarray}
where we have used the EL-EoM: $\,\partial_i H^{0ij} = -\, \big (\partial^0 \bar B^j - \partial^j \bar B^0 \big ) $ which emerges out\footnote{It is worthwhile to point out that we have {\it not} used the EL-EoM: 
$\,\partial_i H^{0ij} = +\, \big (\partial^0  B^j - \partial^j  B^0 \big )$ which is derived 
(see, e.g. [11]) from the Lagrangian density ${\cal L}_{(B)} $ 
because, in the derivation of the Noether anti-BRST charge $Q_{ab}$, {\it this} Lagrangian density 
plays {\it no} role at all. To be precise, 
the Noether anti-BRST charge $Q_{ab}$ is derived from the {\it perfectly} anti-BRST invariant [cf. Eq. (3)] Lagrangian density ${\cal L}_{(\bar B)}$. } from
the Lagrangian density ${\cal L}_{(\bar B)} $ [cf. Eq. (1)]. 
Thus, taking into account our result of equation (25), we have the following expression for the final form of equation (23), namely;
\begin{eqnarray}\label{28}
s_b \,Q_{ab} = \int d^3 x\,\, \Big [\big (\partial^0 B^i - \partial^i B^0 \big ) \, \bar B_i - 
\big (\partial^0 \bar B^i - \partial^i \bar B^0 \big ) \,  B_i
\Big ],
\end{eqnarray}
which can be re-expressed (using the simple algebraic tricks)  as follows:
\begin{eqnarray}\label{29}
&&s_b \,Q_{ab} = \int d^3 x\,\, \Big [\Big (\partial^0 \{B^i + \bar B^i + \partial^i \phi \} - \partial^i 
\{B^0 + \bar B^0 + \partial^0 \phi \} \Big ) \, \bar B_i \nonumber\\
&&- 
\big (\partial^0 \bar B^i - \partial^i \bar B^0 \big ) \,  \big (B_i + \bar B_i + \partial_i \phi \big ) 
+ \big (\partial^0 \bar B^i - \partial^i \bar B^0 \big )\, \partial_i \phi \Big ].
\end{eqnarray}
Taking the help from the partial integration along with the Gauss divergence theorem, the {\it last} integral of the above equation  can be written as 
\begin{eqnarray}\label{30}
\int d^3 x\;
\big (\partial^0 \bar B^i - \partial^i \bar B^0 \big )\, \partial_i \phi = -\, \int d^3 x\;
\partial_i \big (\partial^0 \bar B^i - \partial^i \bar B^0 \big )\,  \phi \equiv  \int d^3 x\;
\partial_i \big (\partial_j H^{0ji} \big )\,  \phi = 0,
\end{eqnarray}
where we have used, once again,  the EL-EoM: $\partial_j H^{0ji} = -\, \big (\partial^0 \bar B^i - \partial^i \bar B^0 \big ) $ which is derived from
the Lagrangian density ${\cal L}_{(\bar B)} $ [cf. Eq. (1)]. Thus, finally, we have obtained 
\begin{eqnarray}\label{31}
s_b Q_{ab} = -i \, \big \{ Q_{ab}, \; Q_b \big \} &\equiv& \int d^3 x\,\, \Big [\Big (\partial^0 \{B^i + \bar B^i + \partial^i \phi \} - \partial^i 
\{B^0 + \bar B^0 + \partial^0 \phi \} \Big ) \, \bar B_i \nonumber\\
&-& 
\big (\partial^0 \bar B^i - \partial^i \bar B^0 \big ) \,  \big (B_i + \bar B_i + \partial_i \phi \big ) \Big ],
\end{eqnarray}
which establishes clearly that the absolute anticommutativity property (i.e. $\{ Q_{ab}, \; Q_b \} = 0 $) between the Noether conserved
anti-BRST and BRST charges can be proven to be {\it true} if and only if we invoke the 
validity of the CF-type restriction: $B_\mu + \bar B_\mu + \partial_\mu \phi = 0 $.

Against the backdrop of the above thorough discussions, we sketch briefly the proof that: $s_{ab} \,Q_b = -\, i\, \{Q_b, \; Q_{ab} \} = 0 $ is
satisfied if and only if we invoke the validity of the CF-type restriction: $B_\mu + \bar B_\mu + \partial_\mu \phi = 0 $. In this
context, first of all, we observe: 
\begin{eqnarray}\label{32}
s_{ab} \,Q_{b} &=& -\, \int d^3 x\,\, \Big [\big (\partial^0 C^i - \partial^i C^0 \big ) \, \partial_i \rho + \big (\partial^0 \bar C^i 
- \partial^i \bar C^0 \big ) \, \partial_i \lambda + \big (\partial^0 \bar B^i - \partial^i \bar B^0 \big ) \,  B_i \nonumber\\
&-& \dfrac{1}{2}\, \lambda \, \dot \rho + \dfrac{1}{2}\, \rho \, \dot \lambda +  H^{0ij} \, \partial_i \bar B_j\Big ]
\equiv -\, i\, \Big \{Q_{b}, \; Q_{ab} \Big \}.
\end{eqnarray}
Using the partial integration along with the Gauss divergence theorem, we note that the above equation can be re-expressed as follows:
\begin{eqnarray}\label{33}
s_{ab} \,Q_{b} &=& +\, \int d^3 x\,\, \Big [\partial_i \big (\partial^0 C^i - \partial^i C^0 \big ) \,  \rho + \partial_i \big (\partial^0 \bar C^i 
- \partial^i \bar C^0 \big ) \,  \lambda - \big (\partial^0 \bar B^i - \partial^i \bar B^0 \big ) \,  B_i \nonumber\\
&+& \dfrac{1}{2}\, \lambda \, \dot \rho - \dfrac{1}{2}\, \rho \, \dot \lambda +  \big (\partial_i H^{0ij} \big) \,  \bar B_j\Big ].
\end{eqnarray}
At this stage, we have to use the EL-EoM (20), (26) and $\partial_\mu H^{\mu\nu\sigma} + (\partial^\nu B^\sigma - \partial^\sigma B^\nu) = 0 $
(see, e.g. [11]), to simplify the above equation into a short form as follows
\begin{eqnarray}\label{34}
s_{ab} \,Q_{b} = +\, \int d^3 x\,\, \Big [\big (\partial^0 B^i - \partial^i B^0 \big ) \, \bar B_i
- \big (\partial^0 \bar B^i - \partial^i \bar B^0 \big ) \,  B_i \Big ],
\end{eqnarray}
where we have used the specific form\footnote{This is {\it essential} because we are dealing 
with the {\it direct} operation of the anti-BRST symmetry
transformation operator $s_{ab}$ on the explicit expression for the BRST charge $Q_b$ which is derived the BRST symmetry invariance of the action integral
corresponding to the  Lagrangian density ${\cal L}_{(B)} $. We note that the 
r.h.s. of the above equation (34) is {\it exactly} like the r.h.s. of our earlier equation (28).} 
of the EL-EoM: $\partial_i H^{0ij} =  (\partial^o B^j - \partial^j B^0)$ that is derived from the 
 Lagrangian density ${\cal L}_{(B)} $  [cf. Eq. (1)].  Using the similar kinds of algebraic tricks as we
have used from equation (24) to equation (31), we find that equation (34) can be re-expressed,  
in terms of CF-type restriction (i.e. $B_\mu + \bar B_\mu + \partial_\mu \phi = 0 $), as
\begin{eqnarray}\label{35}
s_{ab}\, Q_{b} = -i \, \big \{ Q_{b}, \; Q_{ab} \big \} &\equiv& \int d^3 x\,\, 
\Big [\big (\partial^0 B^i - \partial^i  B^0 \big ) \,  \big (B_i + \bar B_i + \partial_i \phi \big ) \nonumber\\
&-& 
\Big (\partial^0 \{B^i + \bar B^i + \partial^i \phi \} - \partial^i 
\{B^0 + \bar B^0 + \partial^0 \phi \} \Big ) \, \bar B_i \Big ],
\end{eqnarray}
which proves that if we demand the absolute anticommutativity (i.e. $\{ Q_{b}, \; Q_{ab}  \}  = 0$) 
between the Noether conserved\footnote{As an additional remark, we
would like to add that the CF-type restriction, derived in equations (31) and (35), can {\it not} be obtained from the relationships: 
$s_{ab} \,Q_B = -\, i\, \{Q_B, \; Q_{ab} \}$  and $s_{ab} \,Q_B = -\, i\, \{Q_B, \; Q_{ab} \}$ because the above CF-type restriction is required in the
proof of the absolute anticommutativity (i.e. $\{s_b, \; s_{ab} \} = 0 $) between 
the infinitesimal, continuous and off-shell 
nilpotent BRST and anti-BRST symmetry transformation operators [cf. Eq. (6)] that are
generated by the Noether conserved BRST and anti-BRST charges
[{\it not} by the {\it modified} versions of the conserved (anti-)BRST charges $Q_{(A)B}$]. Exactly similar kinds of statements can be made in the context 
of the (anti-)co-BRST symmetry transformations  and corresponding Noether conserved (anti-)co-BRST charges $Q_{(a)d}$ and 
their {\it modified} versions $Q_{(A)D}$, too.}
 BRST charge and anti-BRST charge,
we we end-up with the validity of the CF-type restriction:  $B_\mu + \bar B_\mu + \partial_\mu \phi = 0 $. 

We are now in the position to derive, in a concise manner,
 another CF-type restriction: ${\cal B}_\mu + \bar {\cal B}_\mu + \partial_\mu \widetilde \phi = 0 $  from
the requirements of the absolute anticommutativity (i.e. $s_{ad} \,Q_d = -\, i\, \{Q_d, \; Q_{ad} \} = 0 \; 
s_{d} \,Q_{ad}= -\, i\, \{Q_{ad}, \; Q_{d} \} = 0 $) between the Noether conserved co-BRST and anti-co-BRST charges. Toward this goal in mind, first of all,
we concentrate on the relationship: $s_{ad} \,Q_d = -\, i\, \{Q_d, \; Q_{ad} \} $ and explicitly compute the l.h.s. (i.e. $s_{ad} \,Q_d $) from
the {\it direct} application of the 
anti-co-BRST symmetry transformation operator $s_{ad}$ [cf. Eq. (11)] on the Noether
conserved co-BRST charge $Q_d$ [cf. Eq. (13)] which leads to:
\begin{eqnarray}\label{36}
s_{ad} \, Q_{d} &=& +\, \int d^3 x\,\, \Big [\big (\partial^0 \bar {\cal B}^i - \partial^i \bar {\cal B}^0 \big ) \, {\cal  B}_i
-  \varepsilon^{0ijk}  \, B_i\, \partial_j \bar {\cal B}_k 
- \big (\partial^0 C^i - \partial^i C^0 \big ) \,  \partial_i \rho \nonumber\\
&-& \big (\partial^0 \bar C^i 
- \partial^i \bar C^0 \big ) \,  \partial_i \lambda  
- \dfrac{1}{2}\, \lambda \, \dot \rho + \dfrac{1}{2}\, \rho \, \dot \lambda \Big ].
\end{eqnarray}
We focus on the last {\it four} integrals of he above equation and
use the partial integration along with the Gauss divergence theorem to obtain
\begin{eqnarray}\label{37}
\int d^3 x\,\, \Big [\dfrac{1}{2}\, \rho \, \dot \lambda - \dfrac{1}{2}\, \lambda \, \dot \rho +
\partial_i\, \big (\partial^0 C^i - \partial^i C^0 \big ) \, \rho 
+ \partial_i\, \big (\partial^0 \bar C^i  - \partial^i \bar C^0 \big ) \, \lambda  \Big ] = 0,
\end{eqnarray}
where we have (i) used the EL-EoM (20) and (26), and (ii) taken care of the anticommutativity (i.e. $\rho\, \lambda + \lambda\, \rho = 0 $) 
property between $\rho$ and $\lambda$. Thus, after our observation in equation (37), we have the following expression for ($s_{ad}\, Q_{d}$), namely;
\begin{eqnarray}\label{38}
s_{ad} \,Q_{d} = +\, \int d^3 x\,\, \Big [\big (\partial^0 \bar {\cal B}^i - \partial^i \bar {\cal B}^0 \big ) \, {\cal  B}_i
-  \varepsilon^{0ijk}  \, B_i\, \partial_j \bar {\cal B}_k \Big ].
\end{eqnarray}
We concentrate now on the {\it second} integral of the above equation [cf. Eq. (38)]
and exploit the theoretical strength of the partial integration along with the Gauss
divergence theorem. These algebraic exercises lead to the following:
\begin{eqnarray}\label{39}
 -\, \int d^3 x\,\, \varepsilon^{0ijk}  \, B_i\, \partial_j \bar {\cal B}_k  = +\, \int d^3 x\,\, \varepsilon^{0ijk}  \,(\partial_j B_i)\,  \bar {\cal B}_k
 \equiv -\, \int d^3 x\,\, \varepsilon^{0ijk}  \, (\partial_i \,B_j)\,  \bar {\cal B}_k.
\end{eqnarray}
At this juncture, we can use the EL-EoM: $\varepsilon^{0ijk}  \, (\partial_j \,B_k) = (\partial^0 {\cal B}^i - \partial^i  {\cal B}^0) $ 
(see, e.g. [11]) in the above equation 
and substitute it into equation (38) to obtain a useful form, namely; 
\begin{eqnarray}\label{40}
s_{ad}\, Q_{d} = - \, i \,\big \{ Q_d, \; Q_{ad} \big \} \equiv +\, \int d^3 x\,\, \Big [\big (\partial^0 \bar {\cal B}^i - \partial^i \bar {\cal B}^0 \big ) \, {\cal  B}_i 
- \big (\partial^0 {\cal B}^i - \partial^i  {\cal B}^0 \big )  \bar {\cal B}_i \Big ],
\end{eqnarray}
which is similar to the expressions in equation (28) and (31) modulo a sign factor. Using the algebraic tricks that have been exploited {\it between} (i)
equation (24) to equation (31), and (ii) equation (32) to equation (35), it is straightforward to note that we have the following: 
\begin{eqnarray}\label{41}
s_{ad}\, Q_{d} &=& - \, i \,\big \{ Q_d, \; Q_{ad} \big \} \equiv  +\, \int d^3 x\,\, \Big [ 
\big (\partial^0 \{{\cal B}^i + \bar {\cal B}^i + \partial^i \widetilde \phi \}  - \partial^i \{ {\cal B}^0 
+ \bar {\cal B}^0 + \partial^0 \widetilde \phi \} \big )  \bar {\cal B}_i \nonumber\\
&-& \big (\partial^0  {\cal B}^i 
- \partial^i  {\cal B}^0 \big ) \, \big ({\cal  B}_i + \bar {\cal B}_i + \partial_i \widetilde \phi \big )\Big ].
\end{eqnarray}
Without going into the details of mathematical equations and key algebraic tricks, we state here the analogue of the above equation (41) as follows:
\begin{eqnarray}\label{42}
s_{d}\, Q_{ad} &=& - \, i \,\big \{ Q_{ad}, \; Q_{d} \big \} \equiv +\, \int d^3 x\,\, \Big [\big (\partial^0 \bar {\cal B}^i 
- \partial^i \bar {\cal B}^0 \big ) \, \big ({\cal  B}_i + \bar {\cal B}_i + \partial_i \widetilde \phi \big ) \nonumber\\
&-& \big (\partial^0 \{{\cal B}^i + \bar {\cal B}^i + \partial^i \widetilde \phi \}  - \partial^i \{ {\cal B}^0 
+ \bar {\cal B}^0 + \partial^0 \widetilde \phi \} \big )  \bar {\cal B}_i \Big ].
\end{eqnarray}
Thus, it is crystal clear that the requirement of the absolute anticommutativity between the Noether conserved co-BRST and anti-co-BRST charges leads to
the derivation of the (anti-)BRST and (anti-)co-BRST {\it invariant} CF-type restriction: ${\cal B}_\mu + \bar {\cal B}_\mu + \partial_\mu \widetilde \phi = 0$.
This observation is exactly similar to the derivation of the (anti-)co-BRST and (anti-)BRST {\it invariant}
CF-type restriction: $B_\mu + \bar B_\mu + \partial_\mu  \phi = 0$ in the context of the
requirement of the absolute anticommutativity between the Noether conserved BRST and anti-BRST charges.



\section{Conclusions}

In our present endeavor, we have attached a great deal of importance to (i) the off-shell nilpotent (anti-)BRST and (anti-)co-BRST symmetry transformations, 
(ii) the Noether conserved (anti-)BRST and (anti-)co-BRST charges, (iii) the {\it modified} versions of the 
conserved (anti-)BRST and (anti-)co-BRST charges, and
(iv) the derivations of the CF-type restrictions (i.e. $B_\mu + \bar B_\mu + \partial_\mu  \phi = 0, \;
{\cal B}_\mu + \bar {\cal B}_\mu + \partial_\mu \widetilde \phi = 0$) from the requirements of 
the absolute anticommutativity between the Noether conserved (anti-)BRST and
(anti-)co-BRST charges, respectively. We have  
shown that the Noether (anti-)BRST and (anti-)co-BRST charges are {\it not} invariant under the nilpotent
(anti-)BRST and (anti-)co-BRST symmetry transformations, respectively. Hence, {\it these} charges are {\it not} physical quantities from the point of view
of the symmetry considerations. We have derived the {\it modified} versions of {\it these} charges in a consistent manner
which are invariant under the nilpotent (anti-)BRST and
(anti-)co-BRST symmetry transformations. Thus, these {\it modified} charges are {\it physical}. We have established that the Noether
and {\it modified} versions of the conserved charges anticommute with each-other.  Hence, they are linearly {\it independent} of each-other.

We have observed that the Noether conserved (anti-)BRST charges 
$Q_{(a)b}$ are useful (i) as the generators (see, e.g. [13]) for the nilpotent symmetry transformations, and (ii) 
in the derivations of the CF-type restrictions (cf. section four). However, it has been shown, in our earlier work [13], that the {\it modified}
versions of the (anti-)BRST charges $Q_{(A)B}$ are useful in the physicality criteria as they lead to the annihilations of the physical states 
by the operator forms of the first-class constraints [17-19] at the 
{\it quantum} level which are (i) the signatures for  a BRST-quantized gauge theory,
and (ii) consistent with the Dirac quantization conditions for theories that are 
endowed with constraints [18]. In exactly similar fashion, the requirement of the physicality criteria w.r.t.
the {\it modified} versions of the conserved (anti-)co-BRST charges will lead to the annihilation of the physical states (at the {\it quantum} level) 
by the {\it dual} versions of the first-class constraints (see, e.g. [20])
of a BRST-quantized gauge theory (which is a field-theoretic example for Hodge theory).

We  would like to remark that the Lagrangian densities in equation (10) respect {\it four} nilpotent symmetry transformations [cf. Eqs. (2),(11)] because,
under the (anti-)BRST symmetry transformations, we have: $s_{(a)b} [\widetilde \phi, \,{\cal B}_\mu, \,\bar {\cal B}_\mu ] = 0$. Furthermore, we note
that just like equation (6), we {\it also} have: $\{s_d, \; s_{ad} \}\, B_{\mu\nu} = -\, \varepsilon_{\mu\nu\sigma\rho} \, \partial^\sigma ({\cal B}^\rho
+ \bar {\cal B}^\rho) $ which turns out to be {\it zero} due to the validity of 
CF-type restriction: ${\cal B}_\mu + \bar {\cal B}_\mu + \partial_\mu \widetilde \phi = 0$. It would be nice 
future endeavor to extend our present ideas and discussions 
in the context of the 4D BRST-quantized version of the St{\" u}ckelberg-modified {\it massive} Abelian 2-form
gauge theory (see, e.g. [11]) which is {\it also} endowed with a couple of {\it non-trivial} CF-type restrictions.


\begin{thebibliography}{99}
\bibitem{RPM1}    M. B. Green, J. H. Schwarz, E. Witten, Superstring Theory  \\(Cambridge University Press, Cambridge, 1987) 
\bibitem{RPM2}    J. Polchinski, String Theory (Cambridge University Press, Cambridge, 1998)    
\bibitem{RPM3}    D. Lust, S. Theisen, Lectures on String Theory (Springer-Verlag, New York, 1989)    
\bibitem{Hari4}   C. Becchi, A. Rouet, R. Stora, The Abelian Higgs-Kibble model: unitarity of the S-operator. Phys. Lett. B 52, 344 (1974)
\bibitem{SKP5}    C. Becchi, A. Rouet, R. Stora, Renormalization of the Abelian Higgs-Kibble model. Comm. Math. Phys. 42, 127 (1975)
\bibitem{SKP6}    C. Becchi, A. Rouet, R. Stora, Renormalization of gauge theories. \\ Ann. Phys. (N. Y.) 98, 287 (1976)
\bibitem{SKP7}    I. V. Tyutin, Gauge invariance in field theory and statistical physics in operator formalism, in Lebedev Institute Preprint, 
                  Report Number: FIAN-39 (1975) (unpublished), arXiv: 0812.0580 [hep-th]
\bibitem{Hari8}   L. Bonora, R. P. Malik, BRST, anti-BRST and gerbes. Phys. Lett. B 655, 75 (2007)
\bibitem{Hari9}   L. Bonora, R. P. Malik, BRST, anti-BRST and their geometry. \\J. Phys. A: Math. Theor. 43, 375403 (2010)
\bibitem{RPM10}   A. K. Rao, A. Tripathi, B. Chauhan, R. P. Malik, Noether theorem and nilpotency property of the (anti-)BRST charges in 
                  the BRST formalism: a brief review.\\ Universe 8, 566 (2022)
\bibitem{Hari11}  S. Krishna, R. Kumar, R. P. Malik, A massive field-theoretic model for Hodge theory. Ann. Phys. 414, 168087 (2020)
\bibitem{Hari12}  S. Gupta, R. P. Malik, A field-theoretic model for Hodge theory. \\Eur. Phys. J. C 58, 517 (2008) 
\bibitem{RPM13}   R. P. Malik, 	Abelian 2-form gauge theory: basic canonical brackets and nilpotency property of the Noether (anti-)BRST charges,
                  arXiv: 2607.06486 [hep-th]
\bibitem{SKP14}   T. Eguchi, P. B. Gilkey, A. Hanson, Gravitation, gauge theories and differential geometry. Phys. Rep. 66, 213 (1980)
\bibitem{SKP15}   S. Mukhi, N. Mukunda, Introduction to Topology, Differential Geometry and Group Theory for Physicists 
                 (Wiley Eastern Private Limited, New Delhi, 1990)
\bibitem{SKP16}   M. G{\" o}ckeler, T. Sch{\" u}cker, Differential Geometry, Gauge Theories and Gravity\\ 
                 (Cambridge University Press, Cambridge, 1987) 
\bibitem{RPM17} P. A. M. Dirac, Lectures on Quantum Mechanics (Belfer Graduate School of Science) (Yeshiva University Press, New York, 1964)
\bibitem{RPM18} K. Sundermeyer, Constraint Dynamics: Lecture Notes in Physics\\ (Springer-Verlag, Berlin, 1982)
\bibitem{RPM19} E.C.G. Sudarshan, N. Mukunda, Classical Dynamics: A Modern Perspective \\(Wiley, New York, 1972)                
\bibitem{RPM20} R. P. Malik, Nilpotent (anti-)co-BRST symmetries and their consequences in a 4D BRST-quantized field-theoretic system. 
                Nucl. Phys.  B 1025, 117394 (2026)            
\end{thebibliography}
\end{document}